\documentclass[aps,prl,twocolumn,showpacs,a4paper,unsortedaddress,
amsmath,amssymb,byrevtex]{revtex4}

\usepackage[latin1]{inputenc}
\usepackage{graphicx}
\usepackage{dcolumn}
\usepackage{bm}
\usepackage{hyperref}
\usepackage{times}

\begin{document}
\preprint{ffuov/02-01}

\title{Tailoring the Fermi level of the leads in molecular-electronic
devices}

\author{V. M. Garc\'{\i}a-Su\'arez}
\email{v.garcia-suarez@lancaster.ac.uk}
\author{C. J. Lambert}
\affiliation{Department of Physics, Lancaster University,
Lancaster, LA1 4YB, U. K.}

\date{\today}

\begin{abstract}
The dependence of the transport properties on the specific
location of the Fermi level in molecular electronics devices is
studied by using electrodes of different materials. The zero-bias
transport properties are shown to depend dramatically on the
elemental composition of the electrodes, even though the shape of
the transmission coefficients is very similar. By using alkaline
materials it is possible to move the Fermi level from the
HOMO-LUMO gap to the LUMO resonance and change dramatically the
length dependence of the conductance of molecular wires, which
opens the possibility of using molecules with different lengths
and very similar conductances in nanoscale circuits. This method
shows how to dramatically increase the conductance of molecular
devices and alter qualitatively and quantitatively their
electronic and transport properties.
\end{abstract}


\maketitle

\section{Introduction}

During recent years, the race to develop a viable technology for
sub-10 nm electronic wires led a number of groups to study
electron transport through single molecules bridging two metallic
electrodes. Examples include both experimental
\cite{Joa95,Ree97,Dat97,Smi02} and theoretical
\cite{Xue02,Bra02,Fuj03,Cal04,Roc06} works. In all of these
experiments one or more of the metallic electrodes are formed
either from break junctions \cite{Ree97,Smi02} or from the tip of
an STM \cite{Joa95,Dat97}. None of these contacting methods is
capable of being scaled to billions of devices on a single chip.
One route to developing a scalable technology is to first deposit
electrodes with a pre-defined gap and then to deposit molecules
whose length is matched to the electrode gap. Since reproducible
electrode gaps significantly below 10 nm do not currently exist
whereas most studies of single molecules involve molecules of
length $\lesssim$ 3 nm, there is a need for systematic studies of
families of molecules of varying lengths which bridge those length
scales. Current examples include alkanedithiols and
alkanemonothiols \cite{Jia04}, diaminoacenes \cite{Qui07} and
oligothiophenes \cite{Lea08}.

When a low-enough voltage is applied to a single molecule bridging
two metallic electrodes, the measured electrical conductance $G$
is governed by the spatial extent of the molecular orbitals, their
level broadening $\Gamma$ due to the contacts and the position of
the HOMO ($E_\mathrm{H}$) and LUMO ($E_\mathrm{L}$) levels
relative to the Fermi energy ($E_\mathrm{F}$) of the electrodes
\cite{Yal99,Xue01,Sta06}. This dependence is captured by the
Landauer formula, $G(E_\mathrm{F})=\frac{2e^2}{h}
\int_{-\infty}^{\infty}\mathrm{d}E\,T(E)\left( -\frac{\partial
f(E-E_\mathrm{F})}{\partial E}\right)$, where $f(E)$ is the
zero-bias Fermi distribution in the electrodes and $T(E)$ is the
transmission coefficient for electrons of energy $E$ passing from
one electrode to the other. For synthetic molecules such as
oligothiophenes \cite{Lea08} $\Gamma \ll
E_\mathrm{L}-E_\mathrm{H}$ and therefore $T(E)$ exhibits
Breit-Wigner \cite{Bre36} or Fano resonances \cite{Fan61} in the
vicinity of $E_\mathrm{L}$ and $E_\mathrm{H}$. For conjugated
symmetric molecules with extended wave functions along the
backbones, resonant values of $T(E)$ can be of order of unity and
therefore at low-enough temperatures $G(E_\mathrm{F})\sim 2e^2/h$
whenever $E_\mathrm{F}$ aligns with a resonance. In principle,
this high conductance channel could be observable in conjugated
single-molecule wires of arbitrary length, provided inelastic
scattering is negligible. In practice, for most recently-studied
families of molecules the measured conductance is much less than
$2e^2/h$ because $E_\mathrm{F}$ lies in the HOMO-LUMO gap. This
means a serious handicap from the point of view of molecular
circuitry, because the conductance will decrease exponentially as
a function of length, and from the point of view of molecular
sensing, because the movement of resonances produce by other
molecules or environmental changes will have less effect in the
middle of the gap. One approach to overcoming this problem is to
introduce a third gate electrode, which shifts $E_\mathrm{L}$,
$E_\mathrm{H}$ relative to $E_\mathrm{F}$. In what follows we
explore an alternative approach based on varying the electrode
composition to yield a $E_\mathrm{F}$ close to $E_\mathrm{H}$ or
$E_\mathrm{L}$ and consequently high-conductance molecular wires
without the need for a gate electrode.

\begin{figure}
\includegraphics[width=\columnwidth]{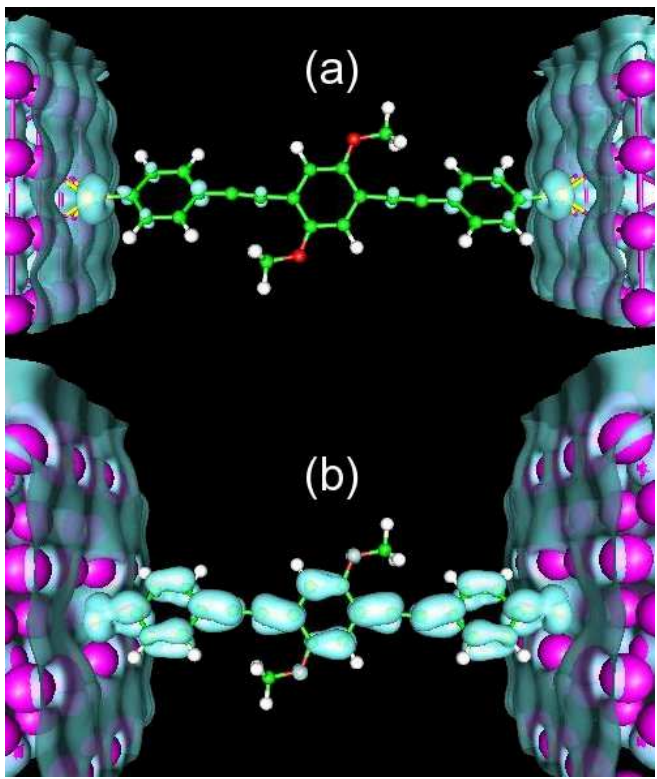}
\caption{\label{Fig1} Real space projection of the density of
states on an interval of 0.2 eV around the Fermi level of the R3
molecule between (a) gold leads and (b) sodium leads.}
\end{figure}

The idea of using alkaline electrodes to tune the work function of
electronic contacts is well known in the organic LEDs (OLEDs)
community \cite{Wak97,Cao00,Brau02}. In this paper, for the first
time, we examine the effect of using alkaline electrodes for
single-molecule electronics. We present a series of ab-initio
calculations based on density functional theory (DFT) \cite{Koh65}
and non-equilibrium Green's functions (NEGF) \cite{Kel65} which
determine the effect of tuning $E_\mathrm{F}$. We examine two
molecules, the archetypal benzene-1,4-dithiol (BDT) molecule that
has been the subject of many experiments and theoretical works
\cite{Ree97,Ven00} (see Fig. (\ref{Fig1})), and a longer molecule
(namely 1,4-bis[4- 24
(acetylsulphanyl)phenylethynyl]-2,6-dimethoxybenzene (molecule R3,
shown in Fig. (\ref{Fig1})) \cite{Coupling} made of three aromatic
rings, with the middle ring slightly twisted by the effect of
oxygen-related side groups. We also examine molecules R5, R7 and
R9 with 5, 7 and 9 aromatic rings, obtained by doubling, tripling
and quadrupling the number of phenyl rings in molecule R3,
respectively.

\section{Theoretical method}

We use the Smeagol code, \cite{Roc06} which interfaces NEGF to the
SIESTA code \cite{Sol02} and obtains self-consistently the density
matrix and the transmission coefficients. We employ a basis set
that includes polarization orbitals (SZP), which is good enough in
this case \cite{SZP}. To approximate the exchange and correlation
we use the local density approximation (LDA) \cite{Per81}, which
works relatively well for organic molecules and noble or alkali
metals. The real-space grid is defined by a plane-wave cutoff of
200 Ry. The leads are made of slices of fcc (gold) or bcc (alkali)
lattices grown along the (001) direction, with 16 atoms per slice
and periodic boundary conditions along the directions
perpendicular to the transport axis, $x$ and $y$. The system is
also made periodic along $z$ to avoid surface effects. The
scattering region includes the molecule and five and six slices of
the electrodes on the left and right parts of the unit cell,
respectively. Each transport calculation has around 250 atoms.

\section{Preliminary calculations}

\begin{figure}
\includegraphics[width=\columnwidth]{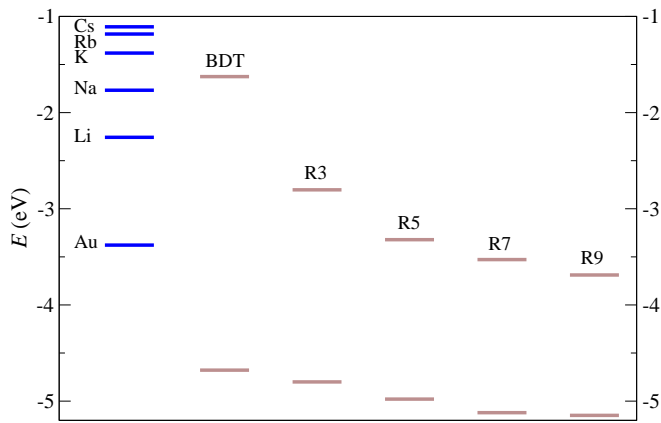}
\caption{\label{Fig2} Relative position of the electrodes Fermi
energies and the molecular HOMO and LUMO states calculated with
LDA and taking as reference the bonding orbital of a hydrogen
molecule away from the system.}
\end{figure}

To evaluate how the Fermi level aligns relative to the molecular
levels we first calculate the energy states of the leads and each
molecule separately. Since in SIESTA the energy origin is
arbitrary \cite{Jun03} it is necessary to use a common reference
to compare the levels of different calculations. For this purpose
we use a hydrogen molecule 10 \AA\ away from the slab or molecule,
whose bonding orbital is taken as reference. The results are shown
in Fig. (\ref{Fig2}). As can be seen, all Fermi levels of the
slabs sits inside the HL gap of the BDT. The Fermi level is
smallest for gold and increases through the alkali metals from Li
to Cs. In the molecules, the HL gap is smaller in R3 than in BDT,
as expected, and decreases as the number of rings increases
towards R9. These results suggest that the conductance of
molecular wires will change dramatically if gold is replaced by
alkali leads. However, they describe only isolated molecules and
do not include bonding to the surface and charge transfer. After
coupling to the slabs the molecular states will broaden into
transmission resonances and the LUMO will shift upwards in energy
towards $E_\mathrm{F}$, due to charge transfer onto the molecule
\cite{Sta06}. To understand these effects we now perform complete
transport calculations.

\begin{figure}
\includegraphics[width=\columnwidth]{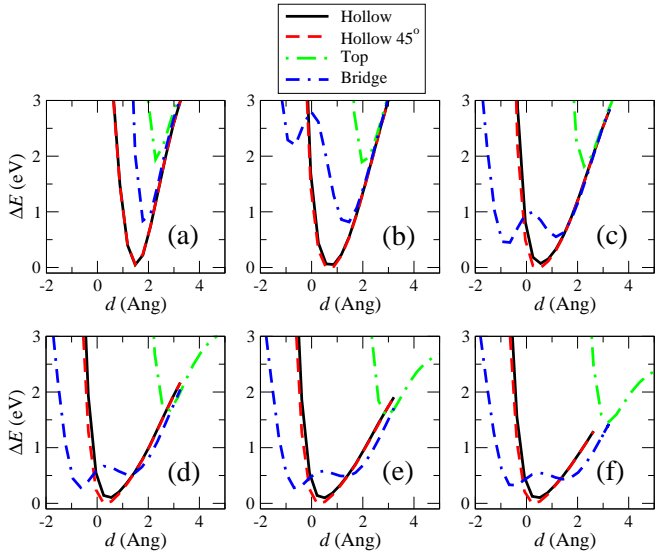}
\caption{\label{Figx1} Energy of a S-C$_6$H$_4$-SH molecule on top
of a surface of fcc gold (a) and bcc lithium (b), sodium (c),
potassium (d), rubidium (e) and cesium (f), as a function of the
distance between the free sulphur and the surface for various
configurations.}
\end{figure}

Since in most of these systems it is not exactly known where the
molecule would attach on the surface, it is necessary first to
determine the most stable bonding configuration. We achieve this
by calculating the energy as a function of the distance between a
S-C$_6$H$_4$-SH molecule and a slab made of 4 layers of fcc Au or
bcc Li, Na, K, R or Cs grown along the 001 direction, as a
function of the distance between the free sulphur atom and the
surface. The results can be seen in Fig. (\ref{Figx1}). The same
calculation was repeated for various positions, which include top,
bridge and a 4-atoms hollow configuration. In the hollow
configuration, steric interactions between some of the atoms of
the molecule and the surface atoms on the corners of the square
can give different energies. We also computed then the case where
the molecule was rotated 45 degrees along the axis perpendicular
to the surface, but the results were practically the same.

We found that in all cases the most stable configuration
corresponds to the hollow position. For short distances, however,
the bridge can become the most favorable bonding configuration in
the case of alkali leads. Its cohesive curve shows also two
minima, as opposed to the one-minimum curves of other
configurations. The second minimum appears when the sulphur atom
crosses the line between both surface atoms and goes deep into the
surface. We expect then that the bonding configuration undergoes a
transition between the hollow and bridge states when the molecule
is compressed. That would also affect the transport properties and
could be detected in the experiments as a small peak in the
conductance histograms. A general trend is that, as the alkali
atomic number increases, the molecule can approach closer to the
surface due to the larger lattice constant of the surface atoms.
This, in turn, slightly increases the bonding distance between the
sulphur and the surface atoms, something which is manifested by
the broadening of the parabola in the cohesion curve from Li to
Cs, and decreases the interaction between the molecule and the
surface.

\section{Alkali vs. gold electrodes}

\begin{table}
\caption{Total charge gained by the molecule after connecting it
to the electrodes obtained from the Mulliken populations.}
\label{Tab01}
\begin{ruledtabular}
\begin{tabular}{lccccccc}
&Au&Li&Na&K&Rb&Cs\\
\hline
$\Delta Q$ (e)&0.45&0.63&0.88&1.15&1.24&1.34\\
\end{tabular}
\end{ruledtabular}
\end{table}

\begin{figure}
\includegraphics[width=\columnwidth]{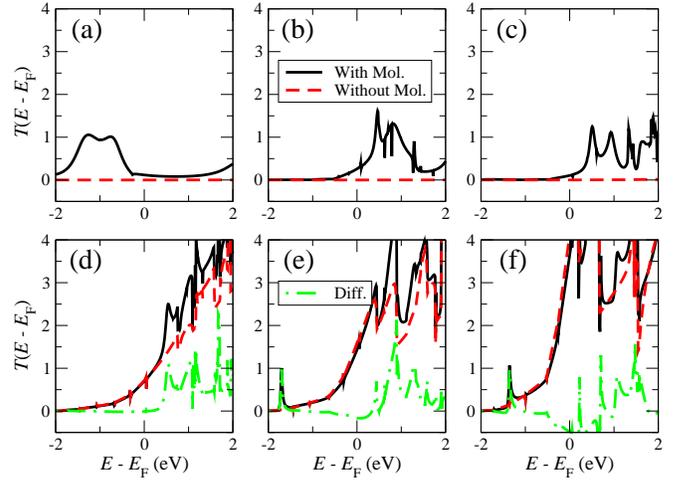}
\caption{\label{Figx2} Transmission coefficients of a BDT molecule
between fcc gold (a) and bcc lithium (b), sodium (c), potassium
(d), rubidium (e) and cesium (f). The dash-dotted line shows the
difference between the transmission curves with and without
molecule.}
\end{figure}

We first simulated the benzene molecule between gold electrodes
and obtained the well known result that the Fermi level is close
to the HOMO resonance \cite{Xue03,Gar07}. When gold was replaced
by Li, however, the LUMO resonance moved towards the Fermi level
and the conductance increased, as can be seen in Fig.
(\ref{Figx2}). This was expected from the results of Fig.
(\ref{Fig2}) due to the change in position of the Fermi level
relative to the molecular orbitals. The same trend was also found
for the other alkali metals. However, for the heavier alkali
elements K, Rb and Cs there was direct transport between the
electrodes even in the absence of the molecule. This effect was
produced by the reduction of the distance between the leads due to
the penetration of the molecule in the surface for large lattice
constants and the increasing electronic delocalization of the
outer electrons as the atomic number increases in these elements.
This can be clearly seen in the lower panels of Fig. (\ref{Figx2})
where the direct transmission increases for high energies, which
correspond to less bounded electrons. In some cases the molecule
acted also as an impurity reducing the total transmission between
the leads. As a consequence, it was not easy to clearly
distinguish the molecular contribution from the electrode
contribution. For that reason, our attention moved to the R3
molecule, for which the electrodes are far enough.

\begin{figure}
\includegraphics[width=\columnwidth]{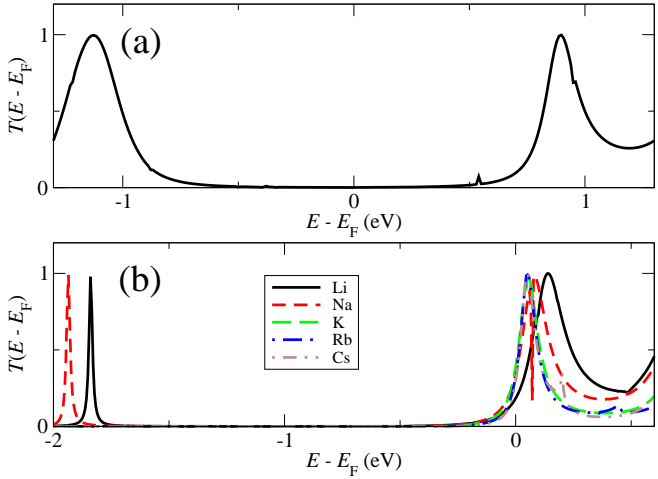}
\caption{\label{Fig3} Transmission coefficients of R3 between gold
electrodes (a) and alkali electrodes (b).}
\end{figure}

Fig. (\ref{Fig3}) shows the transport properties of the R3
molecule between gold and alkali leads. As can be seen, the
transmission coefficients are very similar in all cases and are
characterized by the presence of two main resonances,
corresponding to the HOMO and LUMO orbitals, and a large HL gap.
In the case of gold, the Fermi level sits in the middle of the HL
gap, which produces a very small zero-bias conductance. However,
for the alkali elements the Fermi level moves to the LUMO
resonance and dramatically increases the value of the conductance
\cite{LUMO}. The increase of the conductance can also be seen by
plotting on real space the density of states (DOS) on an energy
interval of 0.2 eV around the Fermi level, as shown in Fig.
(\ref{Fig1}). In the gold case the DOS of the leads does not
penetrate into the molecule and electrons have to tunnel between
both surfaces. When $E_\mathrm{F}$ moves towards the LUMO, DOS
appears also in the molecule and makes it metallic.

We found that the conductance increases from Li to K, but after
that it decreases towards Cs. Such non-trivial behavior can be
explained by taking into account the movement of the Fermi level,
charging effects and changes arising from bonding to different
surfaces. As the alkali atomic number increases, the LUMO
resonance moves downwards (see Fig. (\ref{Fig3})) but as it starts
to cross the Fermi energy, the charge on the molecule increases
and the intralevel repulsion moves the states upwards again
\cite{Sta06}. The total amount of charge transferred to the
molecule increases with the atomic number of the alkali elements,
as can be seen in Table (\ref{Tab01}), which is expected due to
the increasing electropositivity and delocalization. At the same
time, the lattice constant of the contact increases with the
atomic number and therefore the space in the hollow position
becomes larger, which reduces the coupling between the contact
sulphur atom and the surface and decreases the width of the LUMO
resonance. The first effect increases the conductance from Li to
K, whereas the second decreases it towards Cs.

We therefore expect a dramatic increase of the conductance if gold
electrodes were replaced by alkali leads. Based on our
calculations we predict that gold is one of the worst possible
choices for high-conductance molecular nanowires, since the
conductance decreases exponentially as the length of the nanowire
increases.

\begin{figure}
\includegraphics[width=\columnwidth]{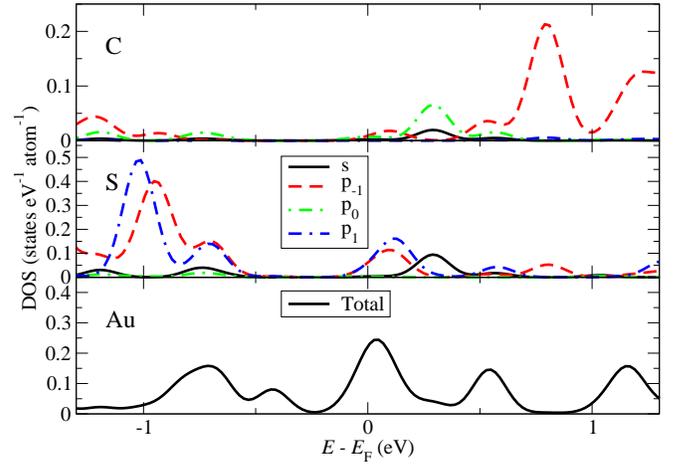}
\caption{\label{Figx4} Projected density of states on the first C
atom, the S atom and the Au electrodes for a R3 molecule in the
same configuration as in Fig. (\ref{Fig3}) (a).}
\end{figure}

From Fig. (\ref{Fig3}) it is possible to deduce some general
properties which can help to understand the influence of various
factors. For example it is noticeable that while the LUMO
resonances in all cases have a similar width, the HOMO resonance
is very sharp when the molecule is coupled to the alkali leads, as
compared to the gold case, and its width decreases with the atomic
number of the alkali element. This arises because the LUMO is
delocalized along the molecule (see Fig. (\ref{Fig1})) and it is
better coupled to the contacts, whereas the HOMO is localized in
the middle of the molecule and is sensitive to any change in the
coupling. Since the coupling to the alkali leads is smaller than
the coupling to gold and decreases from Li to Cs, the width of the
HOMO resonance is much smaller in the former and becomes sharper
as the atomic number increases \cite{HOMORes}.

\begin{figure}
\includegraphics[width=\columnwidth]{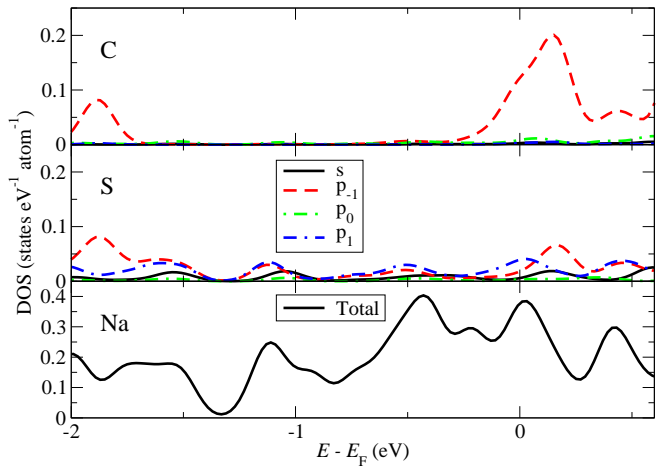}
\caption{\label{Figx5} Projected density of states on the first C
atom, the S atom and the Na electrodes for a R3 molecule in the
same configuration as in Fig. (\ref{Fig3}) (b).}
\end{figure}

Additional information on the electronic structure and its
influence ont the transport properties can be obtained by
analyzing the projected density of states (PDOS), which we show in
Figs. (\ref{Figx4}) and (\ref{Figx5}) for the Au and Na
configurations corresponding to Fig. (\ref{Fig3}). In both cases
the transmission resonances associated with the HOMO and LUMO
orbitals coincide with features in the PDOS of the C and S atoms
related to the p$_{-1}$ orbitals, which are perpendicular to the
transport direction and make on the molecular backbone the
delocalized states near the Fermi level. Such orbitals penetrate
into the molecule and generate the channels associated with the
resonances. In the case of gold there is an additional p$_1$
contribution on the S near the HOMO, due to the strong interaction
with the surface atoms that move these orbitals up in energy, but
it does not go inside the molecule because it almost disappears in
the C atom. The small transmission in the HL gap is produced by
the absence of C states in this region, even though there are S
states generated by hybridization with the surface states. The
density of states of gold and sodium is finite in all the energy
window and is mainly due to the s orbitals. Notice we plot the
density of states per atom, so even in regions where it looks
relatively small it has a finite contribution which allows the
formation of a conductance channel, like for example in Fig.
(\ref{Figx4}) between 0.5 and 1 eV.

\section{Length dependence}

The fact that the Fermi energy is close to the LUMO resonance
opens the possibility of having non-exponential dependence of the
conductance as a function of the molecular length, provided the
Fermi level for longer molecules is still close enough. Such a
possibility would allow the use of molecular wires with various
lengths to connect different parts of a nanoscale circuit with
almost no change in the conductance. We checked this possibility
by using longer molecules with the same central unit and more
benzene rings. To ensure that only the length was varied and no
additional effects were introduced, the rings were made coplanar
with the outer ones and the same number was added on both sides.
The lengths of such molecules, which we call R5, R7 and R9, were
3.4, 4.7 and 6.1 nm, respectively.

\begin{figure}
\includegraphics[width=\columnwidth]{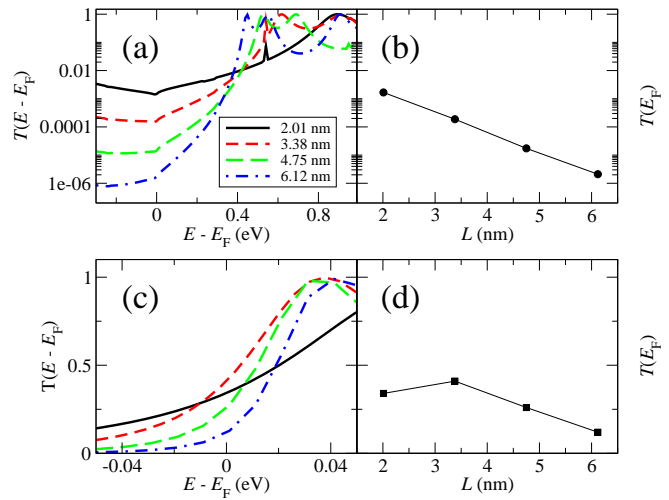}
\caption{\label{Fig4} Length dependence of the transmission
coefficients and low bias conductance for molecules between gold
electrodes ((a) and (b), respectively), and molecules between
sodium electrodes ((c) and (d), respectively).}
\end{figure}

It is well known that as the molecular length increases, the HL
gap decreases due to the higher electronic delocalization along
the molecule. The HL gap and the relative position of the HOMO and
LUMO of all molecules can be seen in Fig. (\ref{Fig2}). However,
as the HL gap decreases, the value of the transmission
coefficients inside the gap decreases due to the larger separation
between the electrodes. This implies again that, if the Fermi
level sits inside the gap, the conductance will decrease
exponentially as a function of the molecular distance. In Fig.
(\ref{Fig4}) we show the transmission coefficients around the
Fermi energy as a function of the distance for gold (a) and sodium
electrodes (c). As can be seen in Figs. (\ref{Fig4})(a) and (b),
which are plotted on a logarithmic scale, the transmission at the
Fermi level decreases exponentially when the molecules are
contacted to gold. Such exponential behavior has a beta of 0.16
\AA$^{-1}$ and indeed proves that the quantum transport is in the
tunnelling regime. As the molecular size increases, the LUMO
resonance moves downwards and the distance between levels
decrease, but these effects are not enough to eliminate the
exponential behavior.

However, in the case of sodium leads, shown in Figs.
(\ref{Fig4})(c), the Fermi level is located near the LUMO
resonance for all molecules. This makes the length dependence
non-exponential and almost linear, as can be seen in Fig.
(\ref{Fig4})(d). Such evolution comes from an interplay between
the downwards shift of the LUMO resonance, charge transfer and
changes in the coupling. The first molecule does not fit well into
the linear trend due to the much bigger width of the LUMO
resonance. Such an anomalous resonance is produced by a very large
interaction between the leads and the molecular central part,
where this orbital has the biggest weight. At the same time, in
order to keep the same charge on the molecule, this orbital is
also shifted upwards. The LUMO resonances of longer molecules have
a width which decreases and a position that increases slightly
with the molecular length. The reduction of the width can be
explained by taking into account that the weight of the LUMO
decreases at the edges of the molecule and therefore the coupling
between this orbital and the contacts tend to decrease as the
length increases (the same, but more pronounced, happens to the
HOMO orbital). The upwards movement of the LUMO resonance looks
counterintuitive, since as the molecular length grows the orbitals
move downwards, but it can be explained by the upwards shift of
the levels produced by the additional charge on the molecule.

\section{Conclusions}

The long-term goal of using high-conductance molecular wires as
interconnects in sub-10 nm electronics requires that HOMO or LUMO
resonances coincide with the Fermi energy of metallic electrodes
so that the conductance does not decrease exponentially and all
nanoscale electronic elements are connected with almost the same
conductance. The pinning of the Fermi level at the LUMO is also
important from the point of view of sensing, since the effect
produced by other molecules or environmental changes, which is
manifested mainly by movements of the resonances, will be much
higher. In this paper we have shown that by analogy with current
technology used in OLEDs, this can be achieved by using alkali
metals \cite{Exp} instead of the more commonly used gold.

\begin{acknowledgments}
We thank Ian Sage for useful discussions. VMGS thanks the EU
network MRTN-CT-2004-504574 for a Marie Curie grant, the EPSRC and
the DTI.
\end{acknowledgments}


\begin{thebibliography}{99}
\bibitem{Coupling}
We have to mention that it is not completely clear that sulphur is
the best atom to couple to alkali leads. A complete study of the
possible coupling atoms is beyond the scope of this work but we do
not expect the main conclusions to be affected by this point
because in general the qualitative trends in the zero-bias
transport characteristics do not depend too much on the coupling
atom \cite{Fin07}.
\bibitem{SZP}
Based on calculations with the BDT molecule, a SZP basis is enough
to accurately describe the results of this work. We note that the
inclusion of polarization orbitals is a key factor to correctly
reproduce the states above the Fermi level, in particular the
LUMO, whereas the inclusion of doube zetas only produces a shift
in the energy of the isolated systems which is corrected when they
become coupled.
\bibitem{LUMO}
We are aware that the position of this orbital is not exactly
given by DFT \cite{Toh05,Koe06}, but it is the trend in the change
of the Fermi level and its movement from the middle of the gap
towards higher conductance values that is of particular interest.
\bibitem{HOMORes}
For K, Rb and Cs the HOMO was so thin that the energy grid used to
plot the transmission coefficients was not enough to represent it.
\bibitem{Exp}
We would like to point out however that from an experimental point
of view the use of alkali electrodes would be a difficult task due
to the high reactivity of such materials.
\end{thebibliography}
\end{document}